# Evolution of Charge and Pair Density Modulations in Overdoped $Bi_2Sr_2CuO_{6+\delta}$

Xintong Li[1], Changwei Zou[1], Ying Ding[2], Hongtao Yan[2], Shusen Ye[1], Haiwei Li[1], Zhenqi Hao[1], Lin Zhao[2], Xingjiang Zhou[2], Yayu Wang[1,3,†]

[1]*State Key Laboratory of Low Dimensional Quantum Physics, Department of Physics, Tsinghua University, Beijing 100084, China*

[2]*Beijing National Laboratory for Condensed Matter Physics, Institute of Physics, Chinese Academy of Sciences, Beijing 100190, P. R. China*

[3]*Frontier Science Center for Quantum Information, Beijing 100084, P. R. China*

[†]Email: yayuwang@tsinghua.edu.cn

One of the central issues concerning the mechanism of high temperature superconductivity in cuprates is the nature of the ubiquitous charge order and its implications to superconductivity. Here we use scanning tunneling microscopy to investigate the evolution of charge order from the optimally doped to strongly overdoped $Bi_2Sr_2CuO_{6+\delta}$ cuprates. We find that with increasing hole concentration, the long-range checkerboard order gradually evolves into short-range glassy patterns consisting of diluted charge puddles. Each charge puddle has a unidirectional nematic internal structure, and exhibits clear pair density modulations as revealed by the spatial variations of superconducting coherence peak and gap depth. Both the charge puddles and the nematicity vanish completely in the strongly overdoped non-superconducting regime, when another type of short-range order with $\sqrt{2} \times \sqrt{2}$ periodicity emerges. These results shed important new lights on the intricate interplay between the intertwined orders and the superconducting phase of cuprates.

# I. INTRODUCTION

A key characteristic of the cuprate high temperature superconductors is the strong tendency for the charge carriers to form symmetry-breaking ordered states [1]. In particular, the charge order phenomena have been found ubiquitously by various experimental techniques in different cuprate families [2-4]. Earlier scanning tunneling microscopy (STM) experiments revealed a checkerboard charge order on the surface of $Bi_2Sr_2CaCu_2O_{8+\delta}$ (Bi-2212) and $Ca_{2-x}Na_xCuO_2Cl_2$ (Na-CCOC), especially when the superconductivity is suppressed by magnetic field, reduced hole density, and elevated temperatures [2,5-7]. More recently, elastic and inelastic x-ray scattering experiments confirmed that the charge order is a bulk phenomenon, and covers a wide range of the cuprate phase diagram [8-10]. Currently, there are still heated debates regarding the nature and implications of the charge order, such as whether it is a nematic order that breaks the $C_4$ rotational symmetry [11-19], and whether it competes with superconductivity [7,14,20,21].

Most previous STM studies on the charge order focused on underdoped and optimally doped cuprates with a well-defined pseudogap phase. It was found that the checkerboard pattern with 4 $a_0$ periodicity ($a_0$ is the lattice constant ~3.8 Å) is the first electronic order that emerges when doping holes into the parent Mott insulator [22], and it persists to hole density close to optimal doping [23]. With further doping into the overdoped regime, resonant x-ray scattering [9] and STM experiments [23] reported the persistence of checkerboard order with increasing wavelength up to 10 $a_0$. However, it remains unknown when and how the checkerboard order eventually vanishes in the strongly overdoped regime. Another important recent progress is the observation that the charge order is closely related to the pair density wave (PDW) order [24-29], i.e., a periodic modulation of the Cooper pairing amplitude. Both scanning Josephson tunneling microscopy on optimally doped Bi-2212 [30] and spectroscopic imaging STM on underdoped Bi-2212 [31] illustrate that the PDW has the same periodicity as the charge order, and they are positively correlated and in phase with each other. However, so far there is no experimental study about the PDW order in the overdoped regime, where superconductivity is severely weakened. Therefore, it is highly desirable to elucidate the doping evolution of both the charge and pair density orders in the overdoped regime of cuprates.

In this article, we report STM studies on $Bi_2Sr_2CuO_{6+\delta}$ (Bi-2201) cuprates with La or Pb substitutions from the optimally doped to strongly overdoped regime. We find that with increasing hole density, the long-range checkerboard charge order gradually evolves into short-range glassy patterns consisting of diluted charge puddles. More interestingly, the unidirectional stripe-like internal structure of the charge puddle pervades in the entire superconducting (SC) phase, and exhibit clear PDW features as revealed by the spatial variations of SC coherence peak and gap depth. Both the charge puddles and the intra-puddle structure vanish completely in the strongly overdoped non-SC regime, when another type of short-range order with $\sqrt{2} \times \sqrt{2}$ periodicity emerges. These results shed important new lights on the intricate interplay between the intertwined orders and the SC phase of cuprates.

High-quality Bi-2201 single crystals with La or Pb substitutions are grown by the traveling solvent floating zone method and post annealed in $O_2$ for extended period, as described in a previous report [32,33]. The hole densities of the four samples studied here are $p = 0.16, 0.19, 0.21$ and $0.23$, which cover the optimally doped to strongly overdoped non-SC regimes. Details about the determination of doping level are discussed in the supplementary section 1. For STM experiments, the crystals are cleaved in ultrahigh vacuum at $T = 77$ K, and then transferred into the STM chamber with the sample stage cooled to $T = 5$ K. The STM topography is taken in the constant current mode with an electrochemically etched tungsten tip calibrated on clean Au(111) surface [34], and d$I$/d$V$ spectra are collected by using standard lock-in technique with a modulation bias voltage with frequency $f = 423$ Hz and 3 mV RMS amplitude.

## II. CHARGE ORDER EVOLUTION

### A. Checkerboard charge order in the optimally doped sample

We first investigate the optimally doped $Bi_2Sr_{1.63}La_{0.37}CuO_{6+\delta}$ with hole density $p = 0.16$ and $T_c = 32.5$ K (denoted as OP-32K). The topographic image in Fig. 1(a) shows that the Bi atoms and the structural super-modulations of the exposed BiO surface can be clearly resolved. Figure 1(b) displays the differential conductance map d$I$/d$V(\boldsymbol{r}, V)$ at a representative bias voltage $V = 10$ mV, which directly visualizes the spatial distribution of local electron density of state (DOS) in this field of view. The most pronounced feature is the well-known checkerboard pattern, in which granular charge puddles with a typical diameter ~ 2 nm form a long-range order along the Cu-O bond direction. These results are highly consistent with previous STM studies on underdoped and optimally doped Bi-2201 [22,35].

To quantitatively characterize the checkerboard order, in Fig. 1(c) we plot the Fourier transform (FT) of the d$I$/d$V$ map in Figs. 1(b), which was widely utilized to determine the modulation wavevector of charge orders in cuprates [5,7,23,36]. The Bragg peaks of the atomic lattice are marked by solid red circles in the FT map, whereas the broad peak circled in blue indicates the wavevector $Q_{CO} \approx 0.21$ ($2\pi/a_0$), corresponding to a wavelength $\lambda_{CO} \approx 4.8\ a_0$ of the long-range checkerboard charge order along the Cu-O direction. The autocorrelation map shown in Fig. 1(d) represents another powerful method to characterize the charge order, as has been demonstrated in Ref. [5], especially if the correlation length of the order is finite. The spatial variations of autocorrelation intensity $AC(\bm{R})$ also reveal the checkerboard order by the four bright spots circled in red, and the white dashed line is a linecut along the Cu-O bond direction. Figure 1(e) displays the $AC(\bm{R})$ linecuts obtained at different bias voltages, where the average distance between neighboring charge puddles, or the charge order wavelength $\lambda_{CO}$, is indicated by the first peak. The position of the peak is non-dispersive from -14 mV to 20 mV, demonstrating the existence of static charge order. The extracted checkerboard wavelength is $\lambda_{CO} \approx 4.7\ a_0$, which is close to $\lambda_{CO} \approx 4.8\ a_0$ obtained by the FT map in Fig. 1(c), as well as that in previous STM work ($\lambda_{CO} \approx 5.0\ a_0$) [35] and resonant inelastic X-ray scattering ($\lambda_{CO} \approx 4.4\ a_0$) [37] on optimally doped Bi-2201.

**B. Charge order evolution in the overdoped regime**

As the hole density is increased to $p = 0.19$, 0.21 and 0.23, $T_c$ is decreased to 24 K (OD-24K), 15 K (OD-15K) and < 2 K (OD-0K), respectively. The charge order phenomena exhibit significant and systematic variations with doping. Figures 2(a) displays the topographic image of the OD-24K sample with chemical formula $Bi_{1.62}Pb_{0.38}Sr_2CuO_{6+\delta}$, in which the surface Bi atoms form a regular square lattice as the structural supermodulation is removed by Pb substitutions [38]. Figures 2(b) is the d$I$/d$V$ map measured at $V = 10$ mV, in which granular charge puddles with typical size ~2 nm are still evident, but the inter-puddle pattern is less close-packed and less ordered than the optimally doped sample shown in Fig. 1(b). This is directly revealed by the FT-map in Fig. 2(c), in which the wavevector corresponding to the checkerboard order becomes barely visible. The $AC(\bm{R})$ map in Fig. 2(d) still has four bright spots (circled in red), but is much weaker and less symmetric than that in Fig. 1(d). By averaging the first peak position along the two directions, the inter-puddle distance is obtained to be $\lambda_{CO} \approx 5.4\ a_0$. The first peak in the $AC(\mathbf{R})$ line-cuts shown in Fig. 2(e) is non-dispersive in the bias range from 0 to 30 mV, so that the observed patterns are also static charge density modulations with short-range correlations.

In the more overdoped OD-15K sample, the checkerboard pattern is further weakened. Figures 3(a) and 3(b) are the topographic image and the d$I$/d$V$ map measured at $V = 10$ mV, in which the granular charge puddles become more diluted and are arranged in a more disordered manner. In the FT-map in Fig. 3(c), the checkboard wavevector totally disappears, and the $AC(\mathbf{R})$ map in Fig. 3(d) only has fuzzy arcs (circled in red). The first peak in the $AC(\mathbf{R})$ line-cuts in Fig. 3(e) is still non-dispersive in the bias range from 0 to 30 mV, and the average distance between the glassy charge puddles is increased to $\lambda_{CO} \approx 6.3\ a_0$.

When the hole density is increased further to $p = 0.23$, the system enters the strongly overdoped non-SC regime. Figures 4(a) and (b) display the topography and d$I$/d$V$ map at $V = 10$ mV for the OD-0K sample, where all the signatures of granular charge puddles ordered along the Cu-O direction disappear. Interestingly, a commensurate $\sqrt{2} \times \sqrt{2}$ charge order along the nearest neighboring Cu-Cu direction emerges, as reported previously [38]. The broken yellow squares in Fig. 4(b) indicate the small areas exhibiting the short-range $\sqrt{2} \times \sqrt{2}$ order. The FT-map in Fig. 4(c) reveals rather weak and broad wavevectors (circled in blue) corresponding to this $\sqrt{2} \times \sqrt{2}$ order, which can be seen more clearly by the cross feature (circled by black dash lines) at the center of the $AC(\mathbf{R})$ map in Fig. 4(d).

The comparison between the FT map and $AC(\mathbf{R})$ map in the overdoped samples clearly demonstrates the advantage of the autocorrelation analysis. When the charge density modulation becomes short-ranged, its feature in the FT map becomes very fuzzy. Instead, the autocorrelation maps are more sensitive to short-range correlations, thus the glassy patterns or small patches of orders can still be revealed.

To quantify the charge puddle density for each sample, we identify the non-dispersive local maxima on d$I$/d$V$ maps as the center of charge puddles (indicated by yellow dots in supplementary Figs. S1(a)-(d)) and count the total number in the measured area. Figure 5(a) indicates the hole concentrations of the samples studies here in the phase diagram, and Fig. 5(b) summarizes the doping dependence of charge puddle density, which decreases continuously with increasing hole density. On the contrary, the $\sqrt{2} \times \sqrt{2}$ charge order that prevails in the OD-0K sample displays an opposite trend. The percentage of the area with $\sqrt{2} \times \sqrt{2}$ charge order (see the estimation in supplementary section 2) grows with overdoing and increases steeply upon entering the non-SC regime. Therefore, in the overdoped regime of Bi-2201, the checkerboard order consisting of granular charge puddles along the Cu-O bond direction becomes more dilute and irregular, and is eventually replaced by the $\sqrt{2} \times \sqrt{2}$ order along the Cu-Cu direction.

**C. The internal structure of charge puddles**

Another intriguing observation is that in the d$I$/d$V$ maps of all three SC samples shown in Figs. 1(b), 2(b), and 3(b), there are bright stripes within each charge puddle pointing to either of the two orthogonal directions. This internal electronic structure is commensurate with the underlying Cu lattice, but breaks the $C_4$ rotational symmetry of the $O_x$-$O_y$ bonds within a $CuO_2$ unit cell. These features are consistent with the nematic order reported before by STM studies in various cuprates [18,39-41]. In the OD-15K sample, the stripy pattern is still so pronounced that it can be directly visualized by the maze-like topography in Fig. 3(a), also similar to that observed previously in lightly doped CCOC [16]. The granular charge puddles and the constituent stripes disappear altogether in the overdoped non-SC sample OD-0K (Fig. 4(b)).

## III. PAIR DENSITY MODULATIONS

**A. Pair density wave in the optimally doped regime**

The evolution of charge order is also closely entangled with the PDW order, which has become a focus topic in the debates of intertwined orders in cuprates [24-31,42,43]. Traditionally, the charge order is revealed by the periodic modulation of d$I$/d$V$ at specific energies, as shown above in the d$I$/d$V$ maps. In order to unveil the PDW order, one must extract the spectral information directly related to the SC properties. In a previous report [31], we have demonstrated that there are two quantities in the d$I$/d$V$ spectrum that directly characterize the strength of local superconductivity. The first is the depth of the SC gap, which reflects the depletion of low energy quasiparticle spectral weight by Cooper pairing. This quantity can be represented by the height difference $H = dI/dV(V_{SC}) - dI/dV(0)$ between the d$I$/d$V$ values at the SC gap edge and zero bias.

The second is the amplitude of the SC coherence peak, which has been shown by various experimental techniques to be related to the local superfluid density [44-49]. This quantity can be characterized by the minus second derivative $D(V) = -d^3I/dV^3$ at the SC gap edge of the d$I$/d$V$ spectrum, which is proportional to the sharpness of the coherence peak. These data analysis methods on Bi-2212 have revealed similar PDW features to that obtained by scanning Josephson tunneling microscopy, which directly probes the local SC order parameter.

We first apply these spectral analyses to the optimally doped Bi-2201 sample. Figure 6(a) displays d$I$/d$V$ curves along a line in OP-32K, which crosses two charge puddles as indicated in the inset. All the spectra possess two separated gaps: a large pseudogap with typical size $\Delta_{PG} \sim 40$ meV and a small SC gap $\Delta_{SC} \sim 10$ meV. Even from the raw data, it can be easily seen that the SC coherence peaks manifest spatial modulations following the charge order, and are much more pronounced on the granular puddles. Three representative d$I$/d$V$ spectra are shown in Fig. 6(b) with the corresponding colored spots marked in the inset. The spectrum taken on top of the bright stripe within a puddle (green curve) has very sharp SC coherence peaks. The blue curve taken on a less bright stripe has relatively weaker coherence peaks, whereas the magenta curve taken between two puddles has shoulders instead of peaks at $V_{SC} = \pm 10$ mV. The depth of the SC gap as represented by the height difference $H = dI/dV(V_{SC}) - dI/dV(0)$ for the three curves are marked by the dashed lines in Fig. 6(b), which clearly reveals that the green, blue and magenta curves have the strongest, intermediate and weakest superconductivity. Figure 6(c) exhibits the $D(V) = -d^3I/dV^3$ curves for the three curves, which directly reveals that the green, blue and magenta curves have the strongest, intermediate and weakest local superfluid density. These two different analysis methods thus give the same conclusion regarding the spatial variations of the strength of superconductivity, although the gap sizes are almost the same.

To directly visualize the spatial distribution of pair density, the SC gap depth $H$ and coherence peak sharpness $D$ are extracted from each d$I$/d$V$ curve acquired on the area in Fig. 1(a), and the $H$-map and $D$-map at $V_{SC} \sim 10$ mV are displayed in Figs. 6(d) and 6(e), respectively. Both images reveal similar long-range checkerboard-like patterns, which unveil the periodic modulation of SC properties, i.e., a well-defined PDW order. More interestingly, in both images the pair density modulation also has the stripe-like internal structure within each puddle, which can already be anticipated from the line-cut in Fig. 6(a). To elucidate the relationship between charge and pair density modulations, the cross correlations of $H(\mathbf{r})$ and $D(\mathbf{r}, V_{SC})$ with the corresponding $dI/dV(\mathbf{r}, V = 10$ mV$)$ are shown in Figs. 6(f) and 6(g). Both cross-correlation maps exhibit a maximum at the center, indicating a positive correlation between the charge and pair density modulations.

## B. Pair density modulations in the overdoped regime

We extend the same measurement and analysis procedures regarding the pair density modulations to the overdoped OD-24K sample with $p = 0.19$. In the inset of Fig. 7(a), a line-cut of $dI/dV$ curves is obtained across a single puddle. For each curve, there are still well-defined SC coherence peaks with gap size $\Delta_{SC} \sim 10$ meV, while the pseudogap size is reduced to $\Delta_{PG} \sim 25$ meV and the features become much weaker. Similar to that in the OP-32K sample, the SC coherence peaks exhibit periodic spatial variations that are strongly correlated to the charge puddle and the stripe-like internal structure. The three representative colored curves in Fig. 7(a) demonstrate that the SC coherence peaks are most pronounced at the bright stripe within a puddle (green curve), become slightly weakened (blue curve) at a less-bright stripe, and evolve into broad shoulders at locations between charge puddles (magenta curve). The $H$-map and $D$-map of the OD-24K sample on the same area as Fig. 2(a) are depicted in Figs. 7(b) and 7(c). Despite the absence of long-range checkerboard order, the periodic distribution of local SC strength still has short-range PDW pattern and pronounced stripe-like internal structure. The insets of Figs. 7(b) and 7(c) illustrate the cross-correlation between the $H$-map and $D$-map with the $dI/dV$ map in Fig. 2(b), and the maximum at the center demonstrates that the pair density modulations are positively correlated with the charge density modulations

When the hole density is increased to $p = 0.21$, the pseudogap of the OD-15K sample becomes too small to be distinguished from the SC gap, as revealed by the $dI/dV$ curves in Fig. 7(d). Nevertheless, the SC coherence peak features still closely follow the granular puddles and stripes, as revealed by the three representative colored curves in the linecut. Even though the long-range inter-puddle ordering disappears, the local puddles and internal stripes are still evident in both the $H$-map (Fig. 7(e)) and $D$-map (Fig. 7(f)). The insets display the cross-correlation maps between them and the charge order map in Fig. 3(b), and the central maximum reveals a positive correlation between the intertwined pair and charge density modulations.

## IV. DISCUSSION

The origin of the various intertwined orders in cuprates, the intricate relationship between them and implications to superconductivity have been core issues in the mechanism problem. Our systematic STM studies on Bi-2201 cover a broad range of phase diagram from the optimally doped to overdoped non-SC regimes (Fig. 5(a)), which was much less explored compared to the underdoped regime. In particular, the quantitative analysis of the overall d$I$/d$V$ lineshape enables us to extract the characteristic features that are selectively sensitive to local superconductivity, and the correlation analysis between various spatial patterns reveal the relationship between the intertwined orders. Our results clarify several important issues concerning the fate of charge and pair density modulations in overdoped cuprates, as will be discussed below.

Firstly, we show that with increasing doping, the long-range checkerboard orders in the underdoped and optimally samples gradually evolve into a glassy state with diluted and disordered charge puddles in the overdoped regime. This picture is distinctively different from the scenario of expanded checkerboard wavelength with overdoping while keeping the ordered structure intact [9,23], despite the deceptive agreement of increased average distance between neighboring puddles in both pictures. The interpretation for the superlattice expansion involves the weak coupling theory of Fermi surface nesting with enlarged hole pocket [4,35], whereas our results indicate that the strong local interaction is still essential even in the overdoped regime.

Secondly, we show that the nematic electronic structure is highly robust against overdoping, and exist in all SC samples studied so far in the Bi-2201 system. As illustrated by the cartoon in Fig. 5(c), the $C_4$-symmetry-breaking stripes form a charge puddle with size ~ 4 $a_0$, or 2 nm, and the puddles are subsequently organized into long-range checkerboard order in the optimally doped sample and diluted glassy state in the overdoped regime (Fig. 5(d)). This observation provides another evidence for the relevance of residual correlation effect between doped holes in overdoped cuprates [11,15,18,50-52].

Thirdly, we show that the charge and pair density modulations are closely entangled, and coexist with the SC phases at least up to $p = 0.21$. More importantly, they vanish altogether in the strongly overdoped limit when superconductivity disappears. This observation suggests that these incipient charge and pair density modulations are not simple competing orders to superconductivity, but may instead share some common origins with Cooper pairing. In the strongly overdoped non-SC regime, they are eventually replaced by the $\sqrt{2} \times \sqrt{2}$ charge order (Fig. 5(e)), which may represent a mechanism for suppressing Cooper pairing [38].

In summary, our STM studies on Bi-2201 cuprates reveal the evolution of charge and pair density modulations in the overdoped regime. We find a gradual deformation of the checkerboard charge order, but the granular puddles with internal nematicity are more robust and exhibit close correlations with local pair density modulations. The dichotomy between the intra-puddle structure and inter-puddle order, plus the complete disappearance of both features in the non-SC regime, shed important new lights on the nature of these intertwined orders in cuprates.

## ACKNOWLEDGMENTS

This work was supported by the NSFC grant No. 11534007, MOST of China grant No. 2017YFA0302900 and No. 2016YFA0300300, and the Strategic Priority Research Program (B) of the Chinese Academy of Sciences (XDB25000000). This work is supported in part by the Beijing Advanced Innovation Center for Future Chip (ICFC).

**Figure Captions**

FIG. 1. The optimally doped Bi-2201 sample with $p = 0.16$. (a) Atomically resolved topographic image acquired at bias voltage $V = -100$ mV and tunneling current $I = 5$ pA over an area of $400 \times 400$ Å$^2$. (b) The $dI/dV$ map measured in the same area as (a) at bias voltage $V = 10$ mV and tunneling current $I = 20$ pA. (c) The Fourier transform map of (b), where the lattice Bragg peaks are marked by red circles and the Cu-O bond direction is indicated by red dashed line. The blue circle highlights the wavevector of the checkerboard charge order. (d) The autocorrelation map of (b) exhibiting a long-range checkerboard charge order, and the four red circles indicate the distance between neighboring charge puddles. The white dashed line indicates the Cu-O bond direction, along which the autocorrelation intensity line-cut is acquired. (e) Bias-dependent autocorrelation intensity $AC(\mathbf{R})$ line-cuts. The red dashed line indicates the position of the charge order peak, which is non-dispersive over a wide range of energies.

FIG. 2. The overdoped OD-24K sample with $p = 0.19$. (a) Atomically resolved topographic image acquired at $V = 100$ mV and $I = 5$ pA over an area of $400 \times 400$ Å$^2$. (b) The d$I/$d$V$ map measured in the same area as (a) at bias voltage $V = 10$ mV. (c) The Fourier transform map of (b), where the blue circle indicates the absence of FT-peak for the checkerboard order. (d) The autocorrelation map of (b), where the short-range charge order is revealed by four bright spots in red circle. (e) Bias-dependent autocorrelation intensity $AC(\mathbf{R})$ line-cuts along the Cu-O bond direction. The red dashed line indicates the non-dispersive charge order peak.

FIG. 3 The overdoped OD-15K sample with $p = 0.21$. (a) Atomically resolved topographic image acquired at $V = -100$ mV and $I = 20$ pA over an area of $400 \times 400$ Å$^2$. (b) The d$I/$d$V$ map measured in the same area as (a) at bias voltage $V = 10$ mV. (c) The Fourier transform map of (b). (d) The autocorrelation map of (b), where the bright arc marked by the red oval indicates the remnant charge order. (e) Bias-dependent autocorrelation intensity $AC(\mathbf{R})$ line-cuts. The non-dispersive first peaks suggest that the fussy arc corresponds to the glassy patterns of charge puddles.

FIG. 4 The overdoped OD-0K sample with $p = 0.23$. (a) Atomically resolved topographic image acquired at $V = 100$ mV and $I = 5$ pA over an area of $350 \times 350$ Å$^2$. (b) The d$I/$d$V$ map measured

in the same area as (a) at bias voltage $V = 10$ mV. (c) The Fourier transform map of (b) displaying the Bragg peaks in the red circles, as well as the short-range $\sqrt{2} \times \sqrt{2}$ charge order in the blue circle. (d) The autocorrelation map of (b), where the central cross and the spot in the circle indicate the short-range $\sqrt{2} \times \sqrt{2}$ charge order in separated patches.

Fig. 5. The doping evolution of charge orders. (a) The schematic electronic phase diagram of Bi-2201, and the arrows indicate the hole densities of the four samples studied in this work. (b) Doping dependence of the charge puddle density and the percentage of areas with $\sqrt{2} \times \sqrt{2}$ charge order. (c)-(e) Schematic cartoons illustrating the evolution of charge order with increasing doping. The yellow spheres represent the charge puddles with size ~4 $a_0$ and the black lines sketch the internal stripy structure. In the overdoped non-SC sample, the charge puddles with stripy internal structure disappear, and small patches with $\sqrt{2} \times \sqrt{2}$ charge order emerge.

Fig. 6. The PDW features in the optimally doped sample. (a) The spectra taken along the yellow arrow in the inset d$I$/d$V$ map. (b) Three representative curves taken on the colored spots in (a), and the gap-depth $H$ of each curve is illustrated by the dashed line. (c) The $D(V) = -d^3I/dV^3$ curves of the three curves in (b). The $D(V_{sc})$ value at $V_{sc}(r) \sim 10$ mV shows the sharpness of local SC coherence peak. (d) The $H$-map exhibits the distribution of local gap-depth $H(r)$. (e) The $D$-map displays the spatial distribution of SC coherence peak sharpness at each location. (f),(g) The intensity distribution of cross-correlation $C(R)$ between the $H$-map and $D$-map with the d$I$/d$V$(10 mV, $r$) map in Fig 1(b). The maximum value at the center indicates a positive correlation.

Fig.7. The pair density modulation features in two overdoped SC samples. (a),(d) Spectral line-cuts along the yellow arrows on the OD-24K and OD-15K samples, respectively. The insets are the corresponding d$I$/d$V$ maps (50 × 50 Å$^2$) taken at bias voltage $V = 10$ mV. (b),(e) The $H$-maps of the OD-24K and OD-15K samples, and the insets are the same cross-correlation plots as in Fig. 6(f). (c),(f) The $D$-maps of the OD-24K and OD-15K samples, and the corresponding cross-correlations are shown in the insets. The maximum at the center indicates a positive correlation between the charge and pair density modulations.

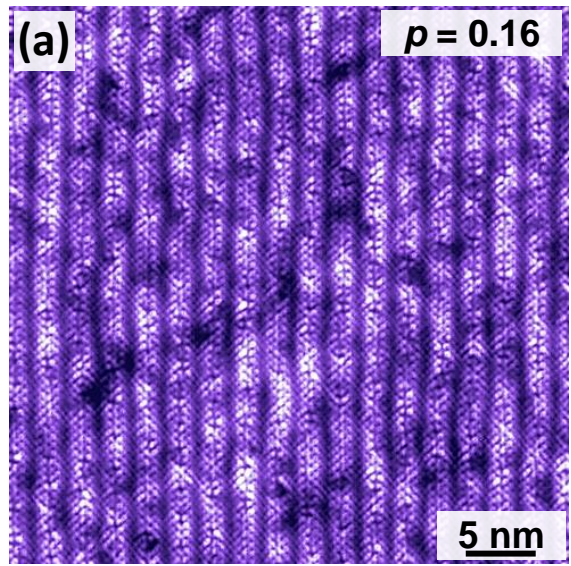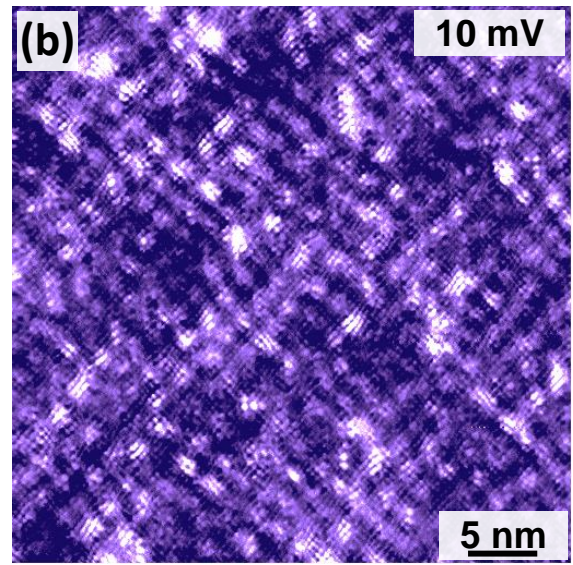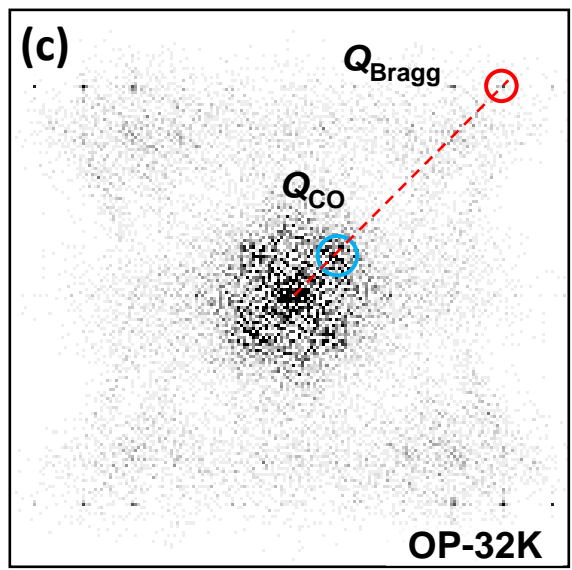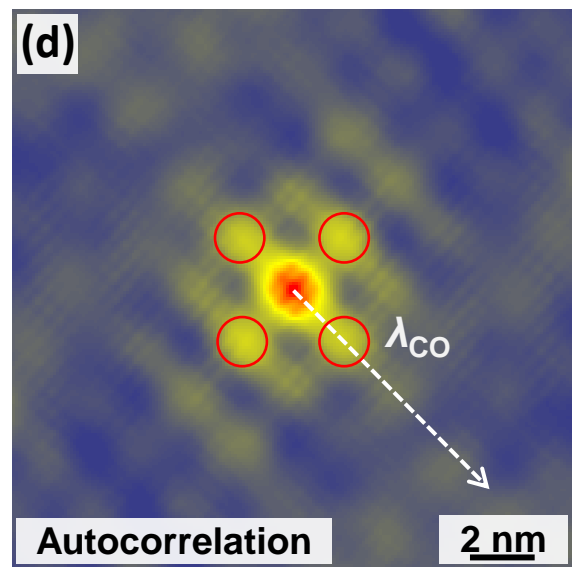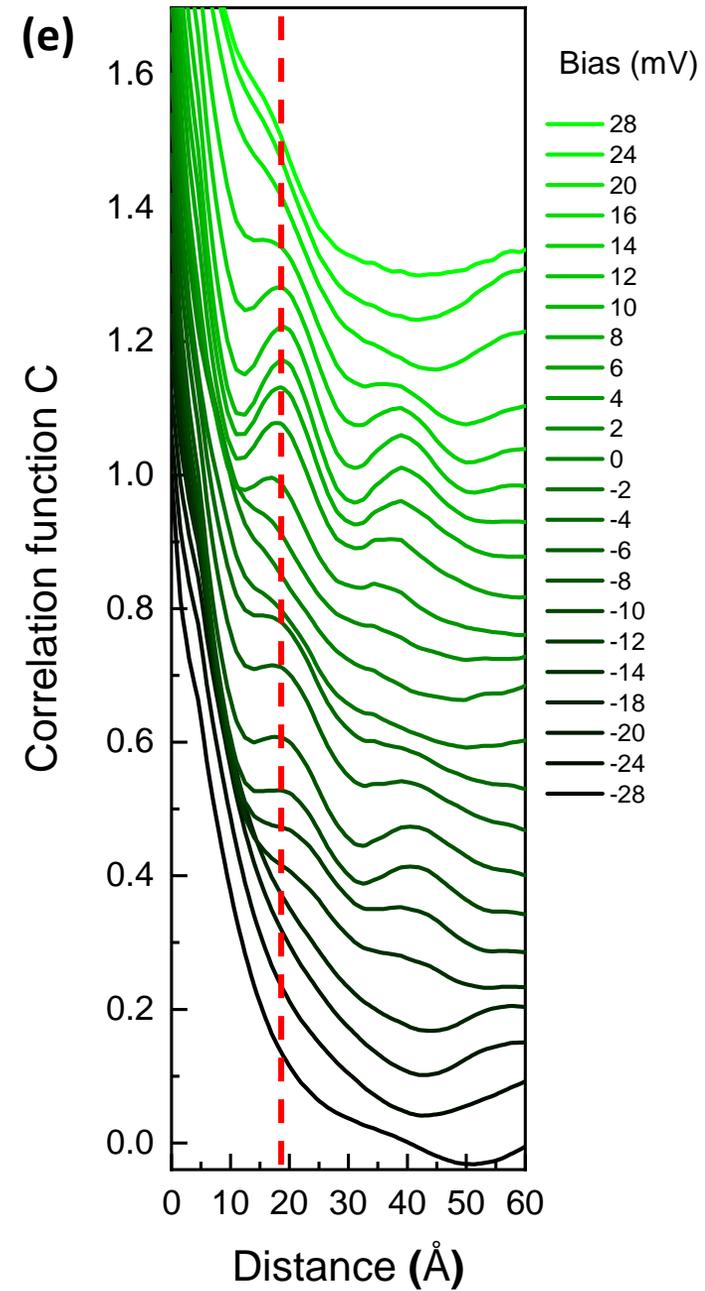

Figure 1

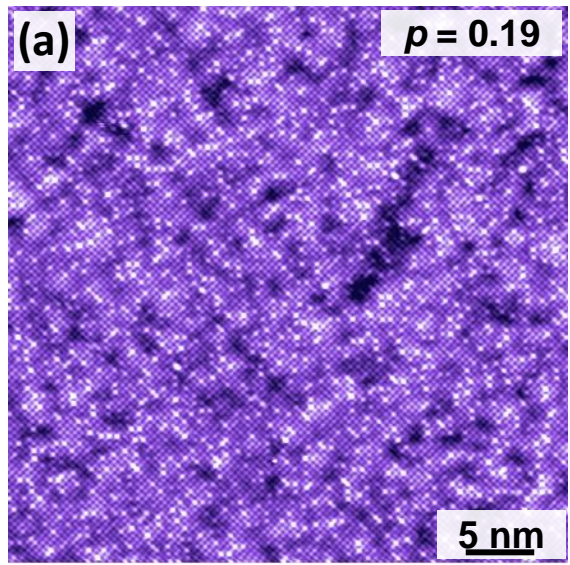
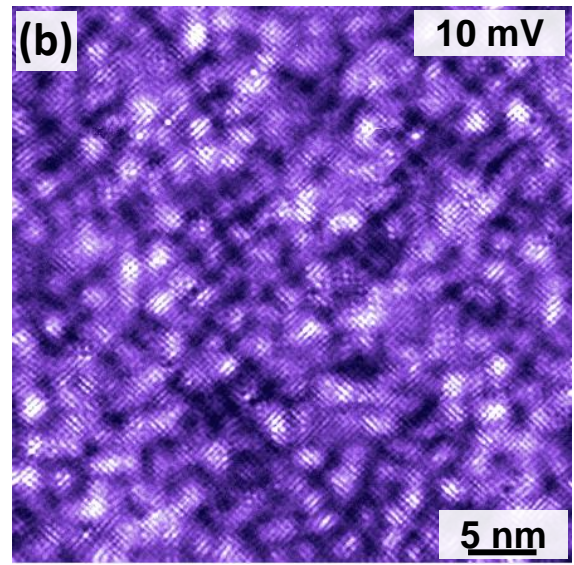
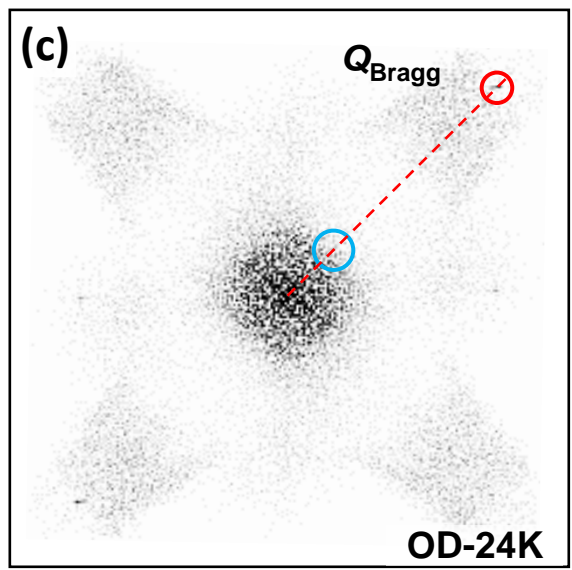
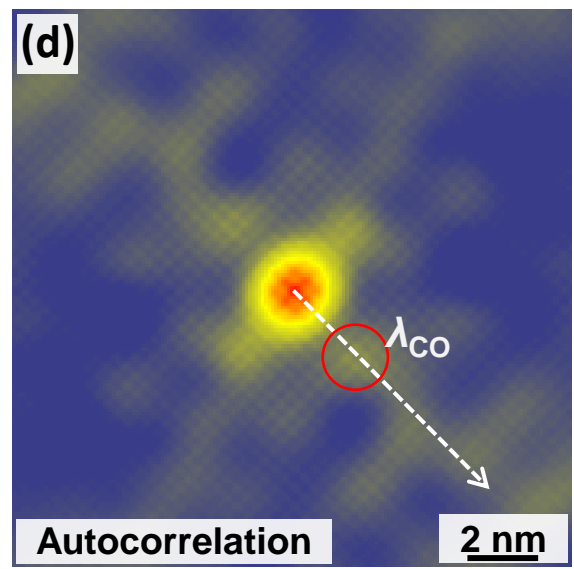
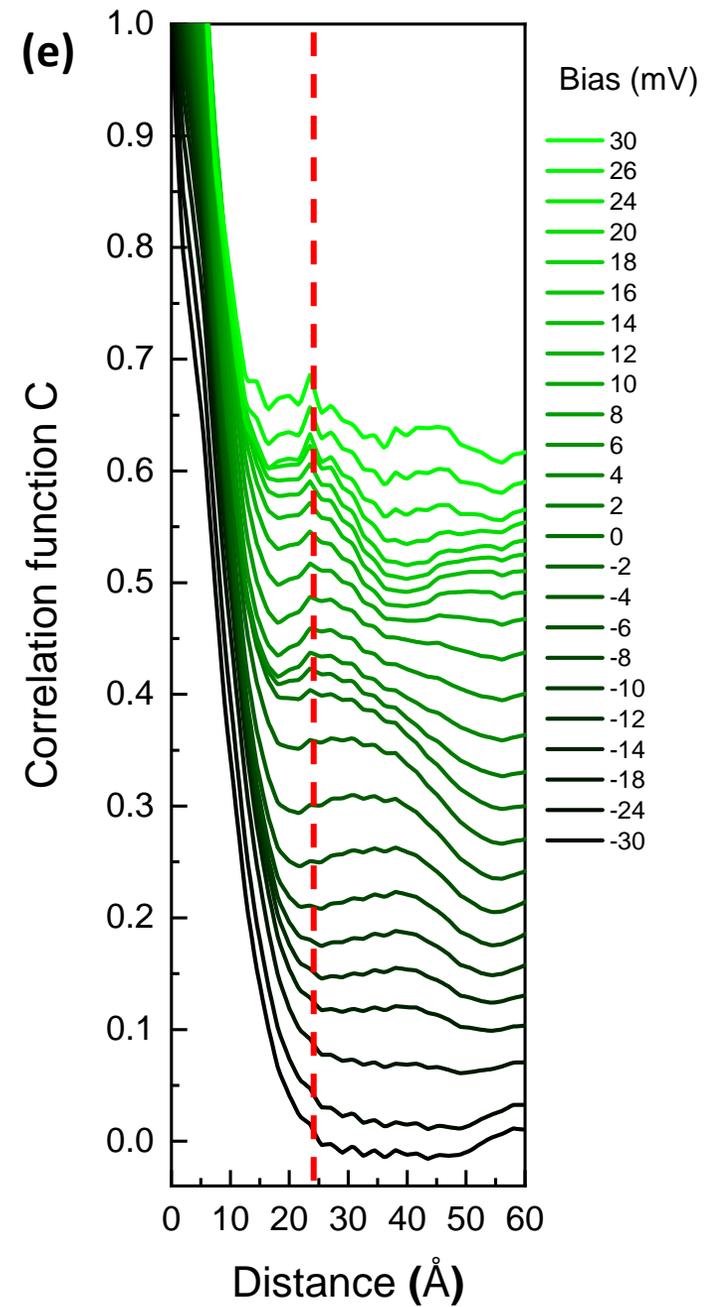

Figure 2

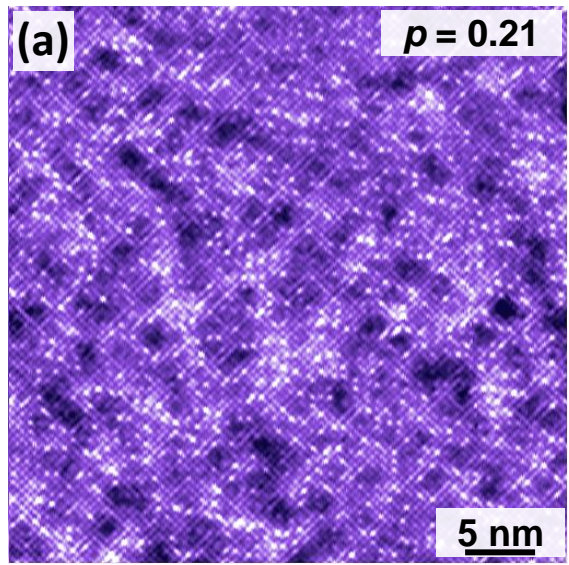 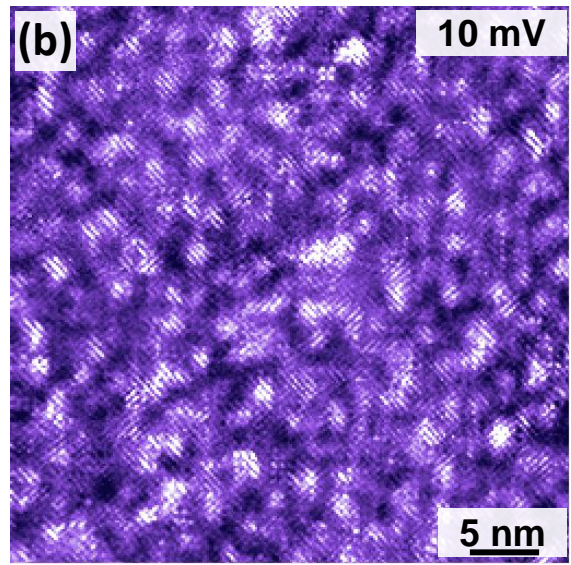
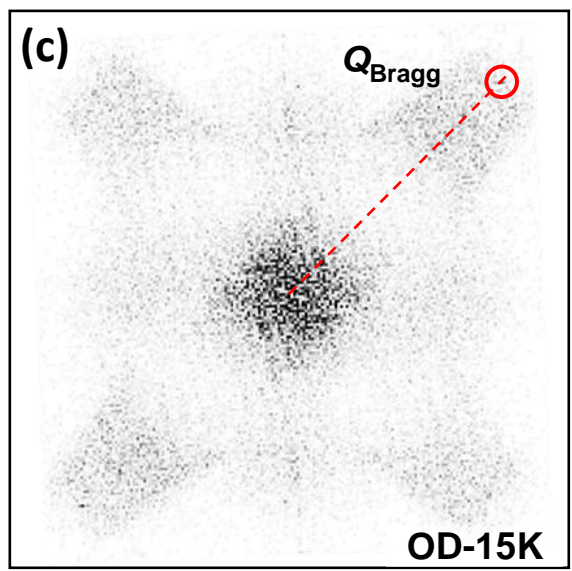 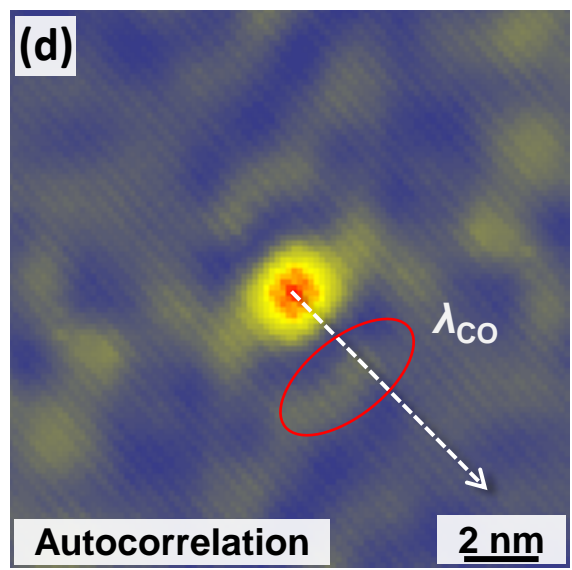
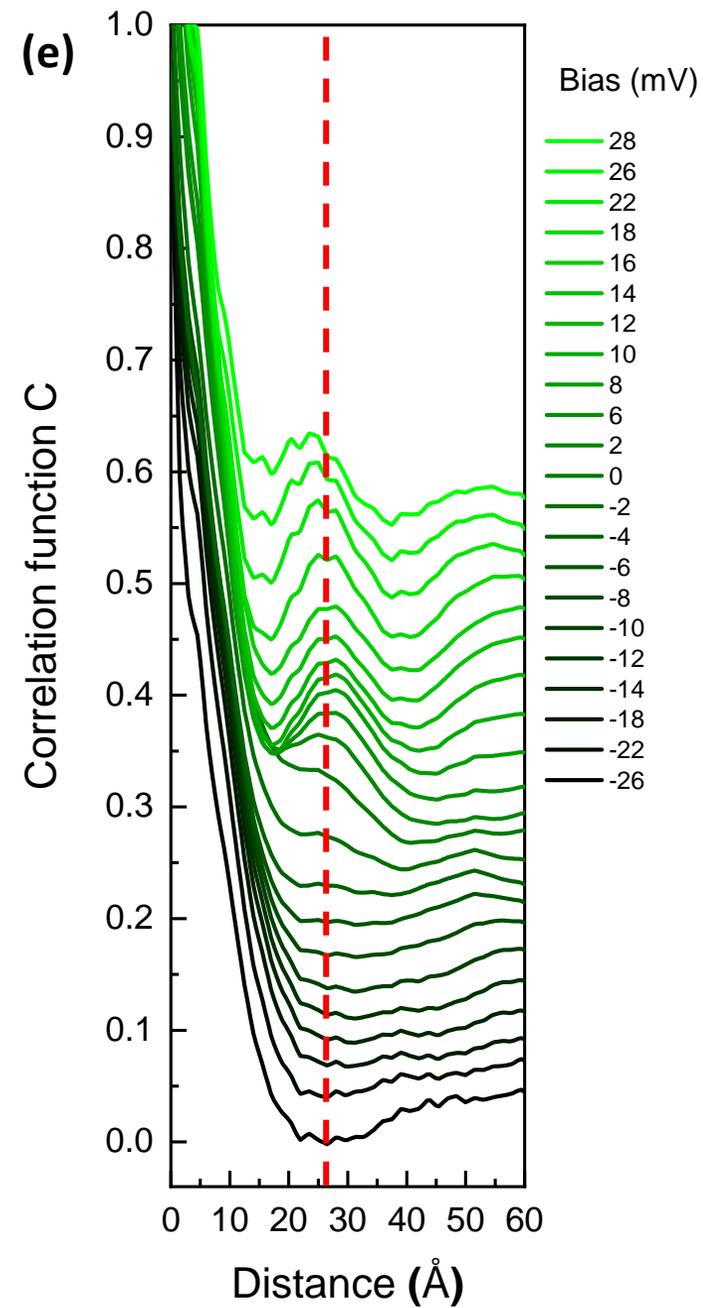

Figure 3

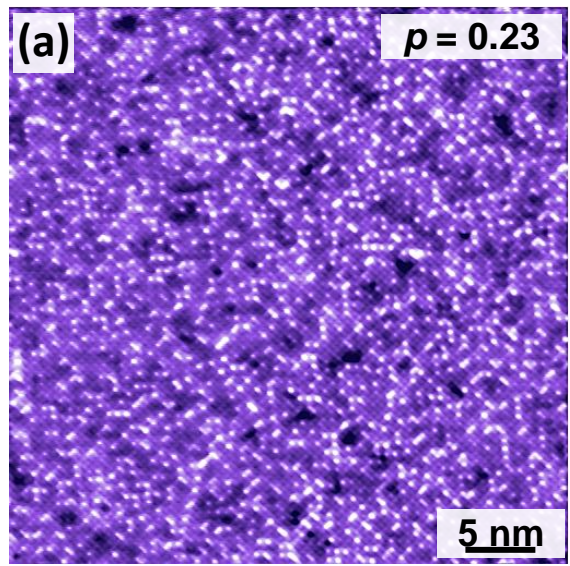 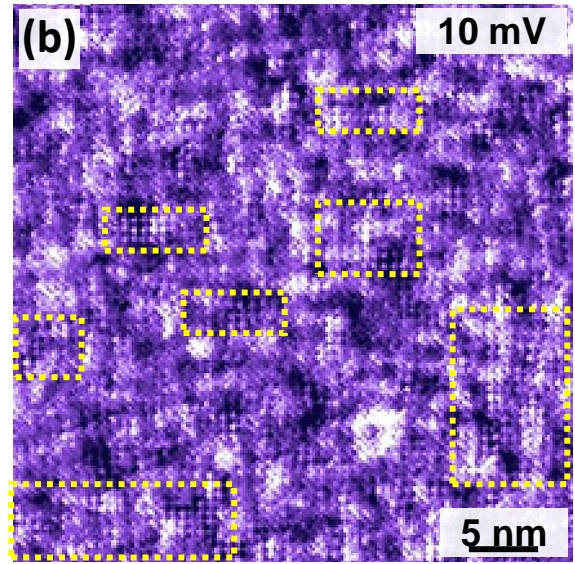 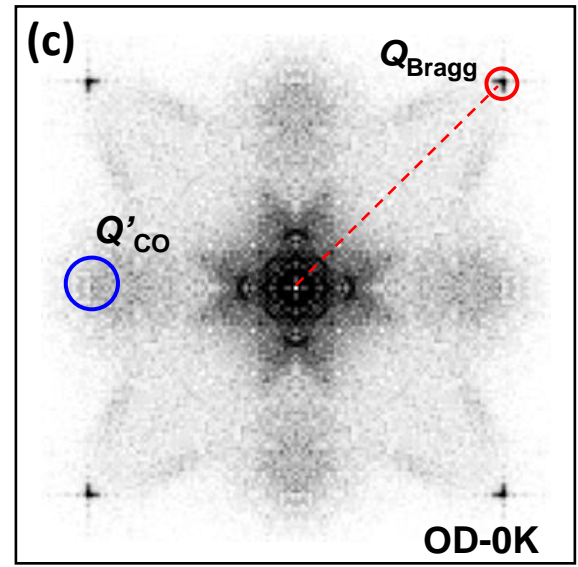 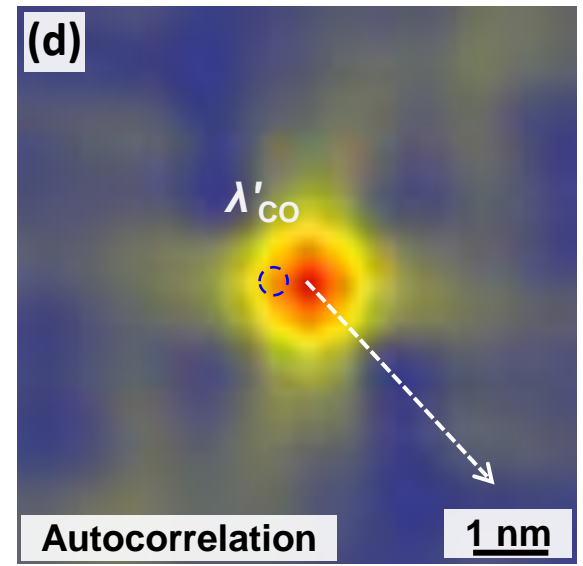

Figure 4

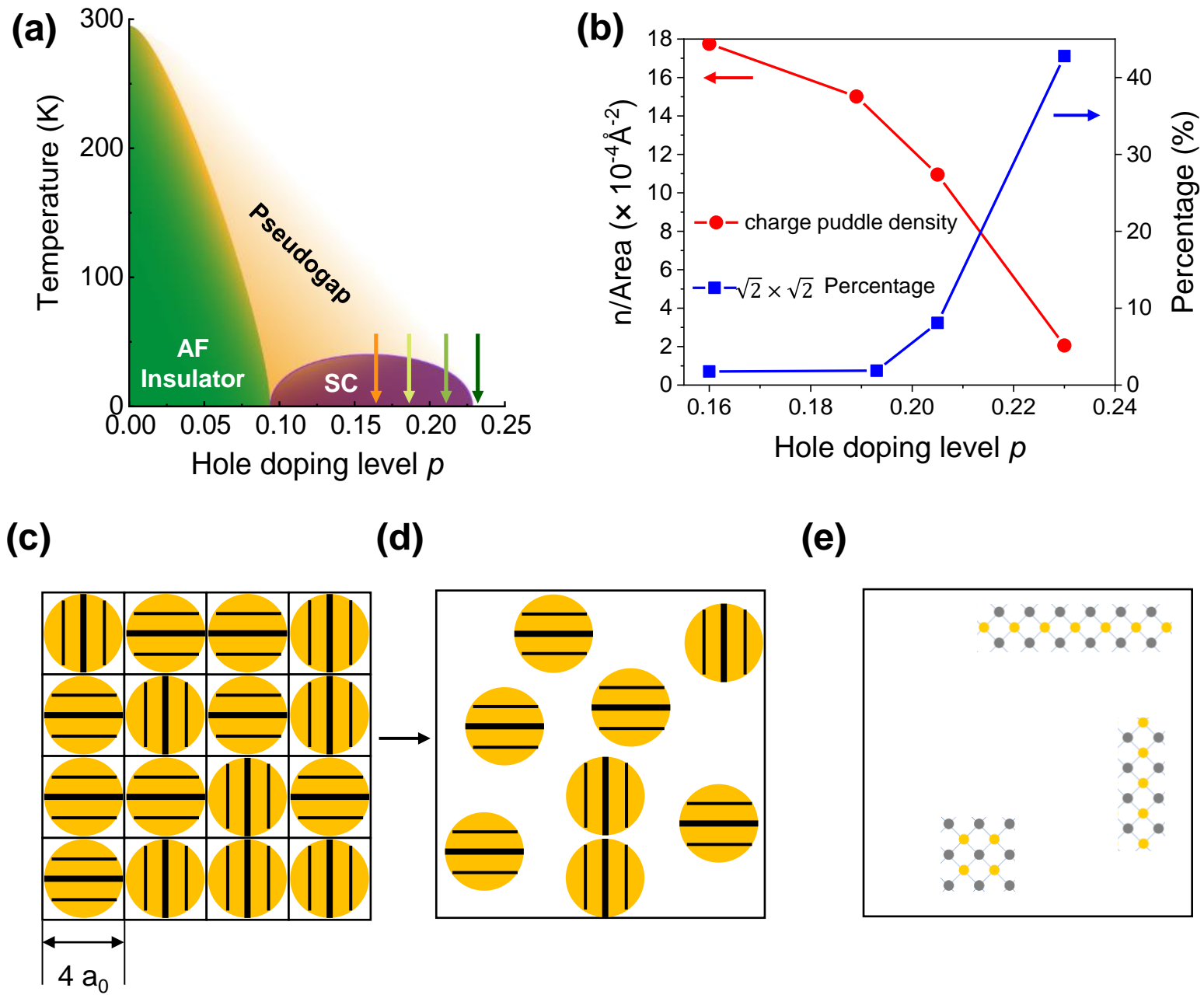

Figure 5

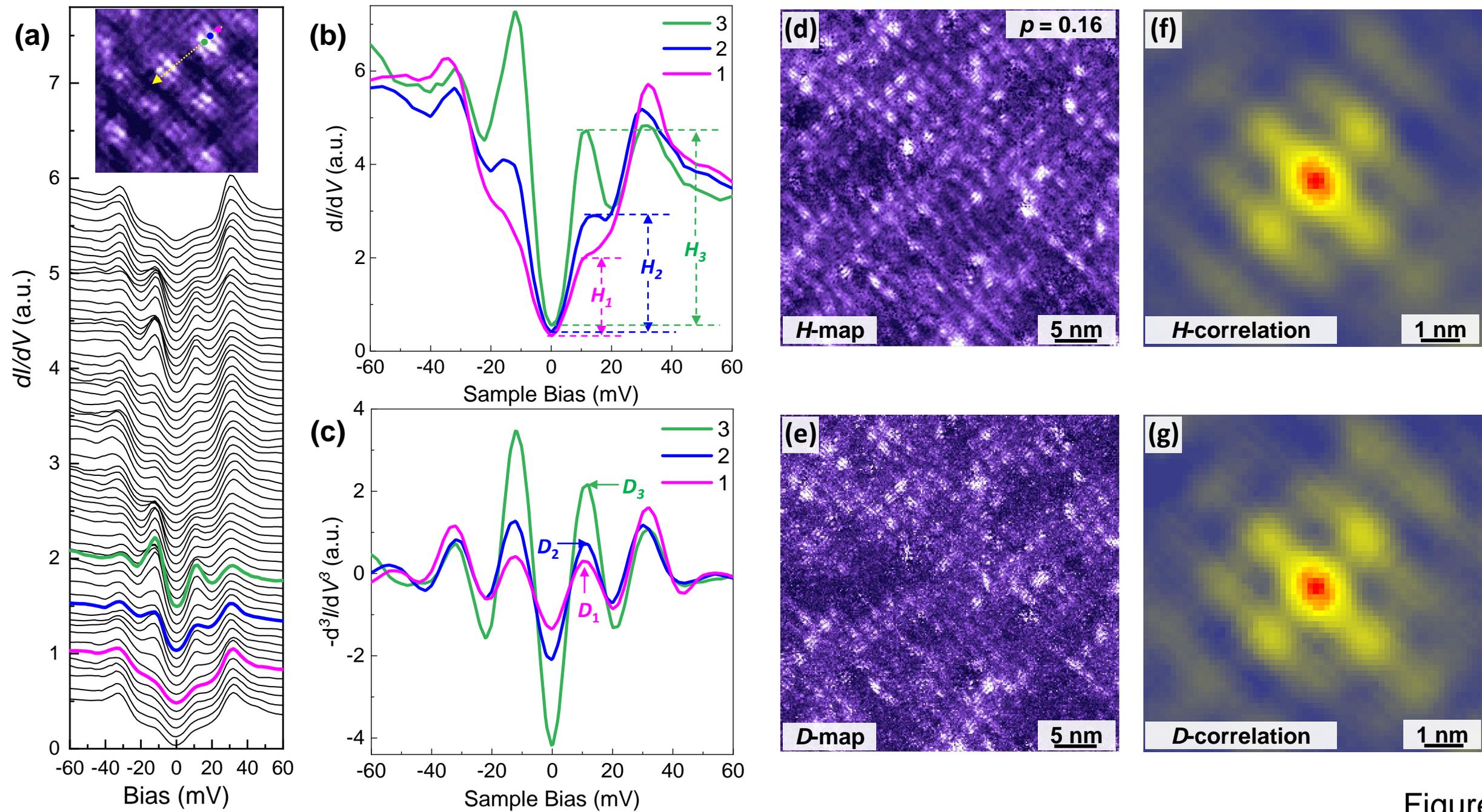

Figure 6

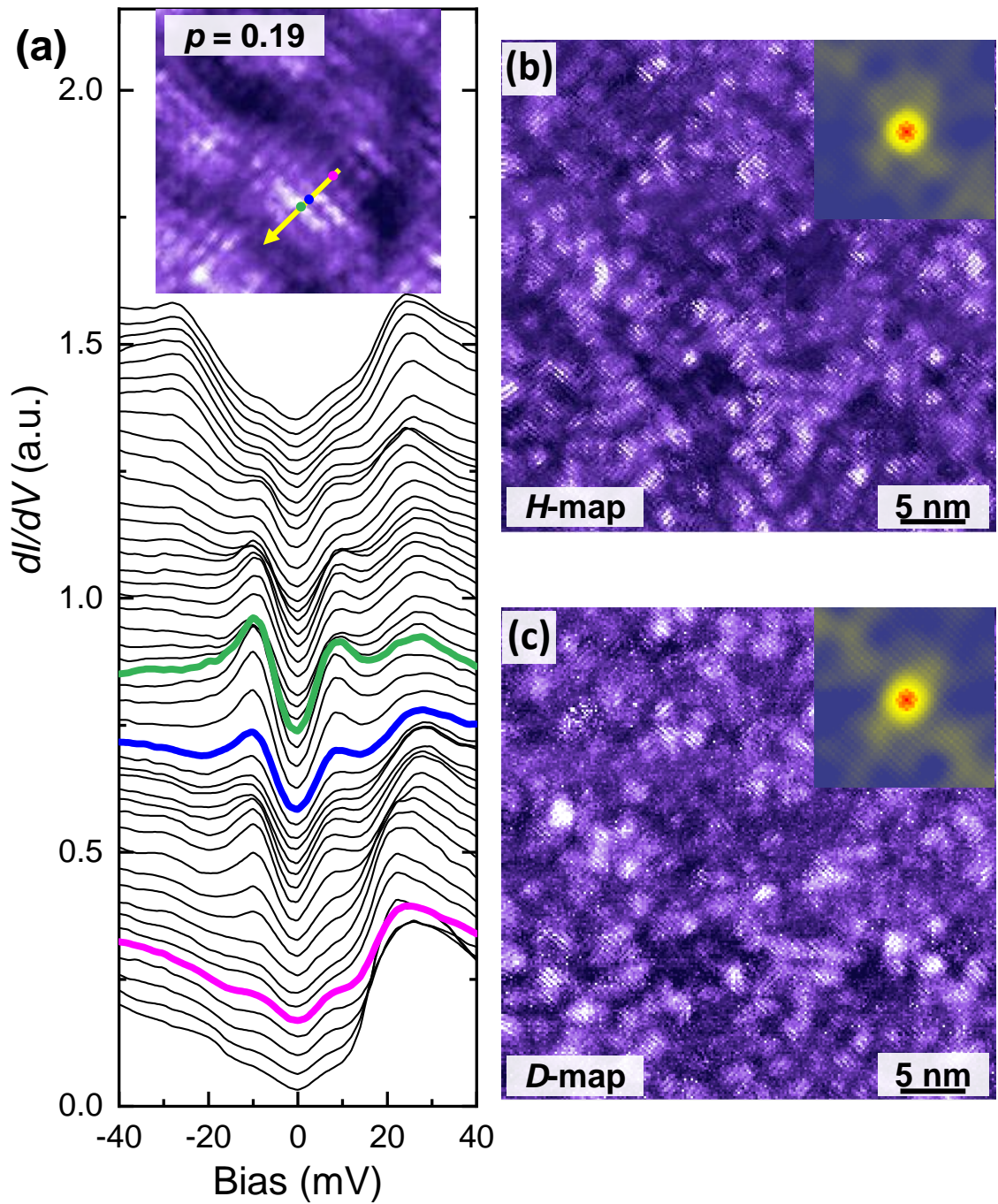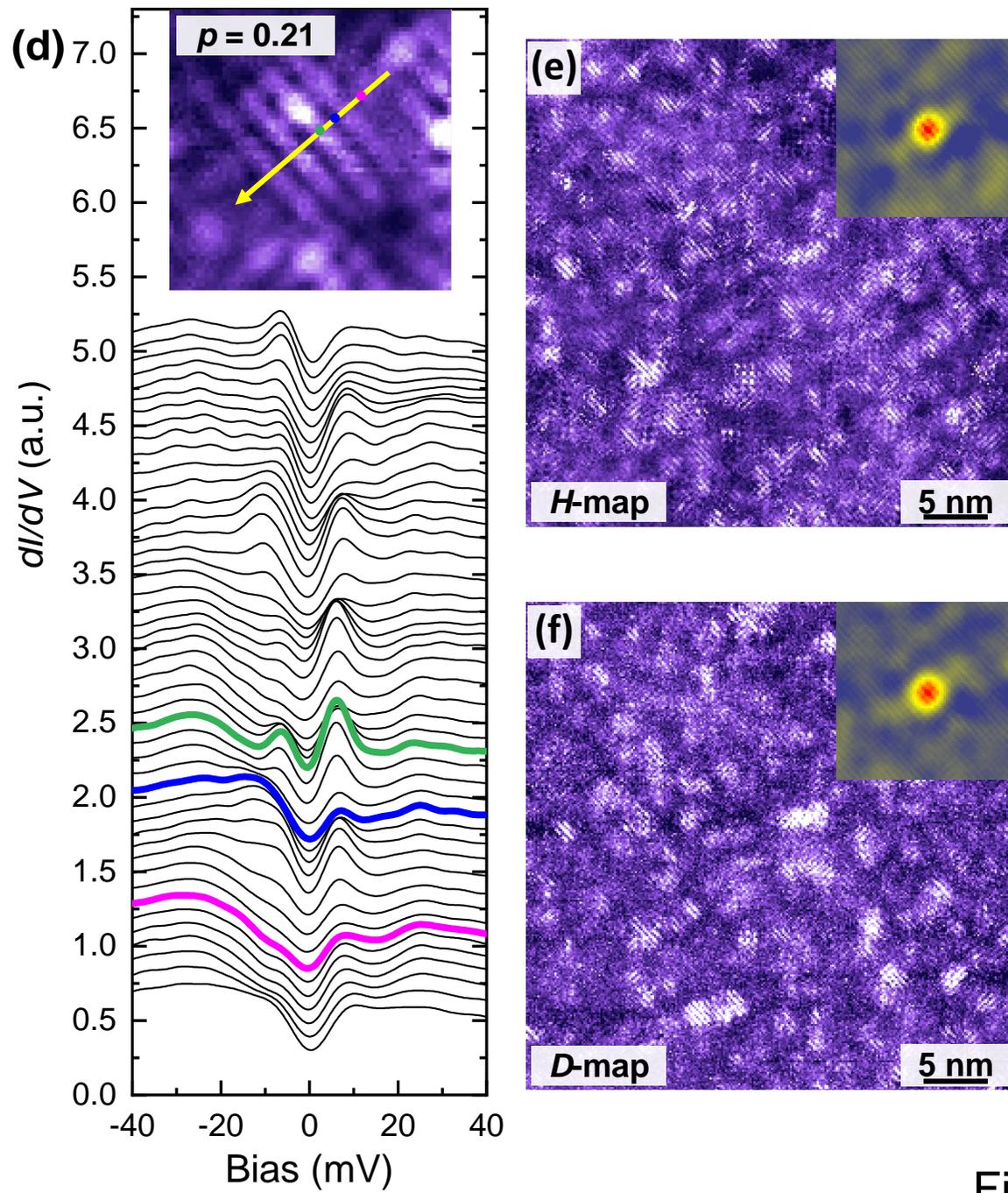

Figure 7